\documentclass[aps,prb,amsmath,amssymb,letterpaper,showpacs,twocolumn]{revtex4}

\usepackage{graphicx}
\usepackage{amsmath}
\usepackage{amssymb}
\usepackage{epstopdf}
\usepackage{bm}
\usepackage{dcolumn}

\providecommand{\braket}[2]{\langle#1\vert#2\rangle}

\providecommand{\ket}[1]{\lvert#1\rangle}
\providecommand{\abs}[1]{\lvert#1\rvert}

\providecommand{\averageop}[3]{\left\langle#1\left\lvert#2\right\rvert#3\right\rangle}

\begin{document}
\title{Tunneling and delocalization effects in hydrogen bonded systems: a study in position and momentum space}
\author{Joseph A. Morrone}
\affiliation{Department of Chemistry \\ Princeton University \\ Princeton, NJ 08544}
\altaffiliation[Present Addresss: ]{Department of Chemistry, Columbia University, New York, NY 10027}
\author{Lin Lin}
\affiliation{Program in Applied and Computational Mathematics, Princeton University \\ Princeton, NJ 08544}

\author{Roberto Car}
\email{rcar@princeton.edu}
\affiliation{Department of Chemistry and Department of Physics, Princeton University \\ Princeton, NJ 08544}

\date{\today}

\begin{abstract}
Novel experimental and computational studies have uncovered the proton momentum distribution in hydrogen bonded systems.  In this work, we utilize recently developed open path integral Car-Parrinello molecular dynamics methodology in order to study the momentum distribution in phases of high pressure ice.  Some of these phases exhibit symmetric hydrogen bonds and quantum tunneling.  We find that the symmetric hydrogen bonded phase possesses a narrowed momentum distribution as compared with a covalently bonded phase, in agreement with recent experimental findings.  The signatures of tunneling that we observe are a narrowed distribution in the low-to-intermediate momentum region, with a tail that extends to match the result of the covalently bonded state.  The transition to tunneling behavior shows similarity to features observed in recent experiments performed on confined water.  We corroborate our ice simulations with a study of a particle in a model one-dimensional double well potential that mimics some of the effects observed in bulk simulations.  The temperature dependence of the momentum distribution in the one-dimensional model allows for the differentiation between ground state and mixed state tunneling effects.
\end{abstract}


\maketitle

\section{Introduction} \label{sec:intro}
The nature of the hydrogen bond plays a critical role in determining the behavior of biological and chemical systems.  The state of a proton that participates in hydrogen bonding, therefore, is a subject of great interest.  This is often characterized in terms of position distributions such as the radial distribution function. The radial distribution function quantifies the probability of one atom being located a certain distance from another atom in position space.  This quantity may be extracted from elastic neutron and x-ray scattering experiments (see e.g. References \onlinecite{eisenberg,soper_2000,headgordon3}), and may also be computed via molecular simulation~\cite{allen}.

Recently, the position space picture that is provided by the radial distribution function has been complemented by measurements of the proton in momentum space.  Deep inelastic neutron Compton scattering experiments~\cite{mayers,noreland,andreani} have uncovered this property in a variety of hydrogen bonded systems, including several phases of bulk water~\cite{reiter,supercool,newexp}, water confined in nanomaterials~\cite{reiternano,reitersilica} and biological systems~\cite{proteinpdist}, ferroelectrics~\cite{kdp,dkdp} and superprotonic conductors~\cite{homouz}.  In these experiments, the observed momentum distribution sharply deviates from classical behavior, underlining the importance of nuclear quantum effects.  These phenomena may be understood in terms of the intertwining of the momentum distribution with the potential energy surface that originates in quantum mechanics due to the uncertanty relation between position and momentum.  As an example of this property, consider the weakening of the oxygen-hydrogen covalent bond as the strength and stability of hydrogen bonding increases. Such effects are typically associated with the red-shift of the OH stretch in the infrared spectra, and may be observed in  the shortening of the tail of the proton momentum distribution~\cite{reiter}.  In certain systems, the proton may be shared equally between the donor and recipient oxygen, thereby forming a so-called ``symmetric'' hydrogen bond.  If one follows the logic of the weakening covalent bond as the hydrogen bond strengthens, then one would expect an increased narrowing of the tail of the momentum distribution as two weak ``symmetric'' bonds are formed between the hydrogen and its neighboring oxygens.  This is precisely the type of behavior that has been observed in experiments on water confined in carbon nanotubes~\cite{reiternano} and the superprotonic conductor Rb$_3$H(SO$_4$)$_2$~\cite{homouz}.

As nuclear quantum effects are essential in determining the form of the momentum distribution, it would seem natural that tunneling phenomena may be studied by means of this property.  Indeed, neutron Compton scattering experiments have presented momentum distributions that report signatures of tunneling in systems such as the ferrolectric potassium dihydrogen phosphate (KDP)~\cite{kdp}, supercooled water~\cite{supercool}, water confined in silica nanopores~\cite{reitersilica} and the hydration shell of globular proteins~\cite{proteinpdist}.  In these systems, a secondary maximum or shoulder is observed that has been related to a node in the tunneling direction of the momentum distribution.  This feature is further supported by a simple analytical model of the ground state of the tunneling wavefunction~\cite{reiter}.  Furthermore, in the experiments that involve supercooled and confined water~\cite{supercool,reitersilica,proteinpdist}, an excess of kinetic energy is observed when compared to liquid water at ambient conditions.  This is manifested in momentum distributions that possess higher density at larger momenta.

The proton momentum distribution may also be accessed via computer simulation. As discussed above, nuclear quantum effects are essential to the computation of this property.  Typical atomic simulations treat the nuclei as classical point particles.  In such simulations the momentum distribution is given by the Boltzmann distribution and exhibits no dependence upon the potential energy surface.  Nuclear quantum effects may be included within the Feynman path integral formulation of quantum statistical mechanics~\cite{feynman}.  In the discretized implementation of this method, one quantum mechanical particle is mapped onto a number of classical replicas that interact harmonically with neighboring replicas, thereby forming a ``chain'' of ``beads''~\cite{chandler}.  The momentum distribution may be computed by means of an ``open'' chain, whereas position-dependent equilibrium properties are computed via ``closed'' chains.  Although the vast majority of path integral simulations solely utilize ``closed'' paths, the momentum distribution has been computed by means of open path integral simulation in superfluid helium~\cite{ceperley2}, and more recently in water~\cite{burnham,morrone,morrone2,burnham08,newexp}.  The latter results complement the  measured experimental proton mometum distributions of water in the liquid and hexagonal solid phases~\cite{reiter,newexp}. 

The open path integral methodology has been recently implemented in conjunction with first principles molecular dynamics~\cite{morrone2} within the Car-Parrinello framework~\cite{cp1}. Previous computational studies of the momentum distribution have not focused on the phenomena of proton tunneling or symmetric hydrogen bonds.  First principles potentials facilitate the study of bond forming and breaking events and thus make the study of these phenomena possible.  Presently, we explore this topic by means of first principles open path integral molecular dynamics studies of high pressure ice.  Although no neutron Compton scattering experiments have been performed on this system, there are several factors that make its study appealing.  High pressure ice has been extensively studied via first principles molecular dynamics simulations~\cite{lee2,lee1,hpice93,hpice94,hpice98,icecpmd,marxice2,hpice02,hpice04,hpice05,benoit05}.  In this work, we will consider three phases of high pressure ice, Ice VII, Ice VIII, and Ice X. By varying the volume of the simulation cells, interconversion of the phases may be induced.  It has been previously shown in the pioneering work of Benoit and Marx~\cite{icecpmd,marxice2} that quantum delocalization and tunneling play a crucial role in the transition between these phases.  Each phase embodies a different hydrogen bonding ``state,'' namely the typical case where the proton experiences relatively strong covalent and weak hydrogen bonding (Ice VIII), proton tunneling along the hydrogen bond from potential well to well (Ice VII), and the ``symmetric'' hydrogen bond (Ice X).  Therefore, its position space distributions are already well characterized in the literature.  Furthermore, since bulk water is particularly amenable to previously developed open path integral methodologies,~\cite{morrone,morrone2} the problem is computationally tractable and several phases may be readily studied.

We find that the uncertainty relation between the position and momentum distributions of the proton~\cite{morrone2} is evident in the behavior of each system.  In particular, the more delocalized protons of Ice VII and Ice X possess more localized momentum distributions in the hydrogen bonding direction.  In Ice X, the symmetric hydrogen bond yields signatures of the momentum distribution that resembles those observed experimentally in other symmetrically hydrogen bonded systems~\cite{reiternano,homouz}.  In the tunneling case, we observe a narrowed momentum distribution with an anomalous shape, namely one that is narrowed at low-momentum, but features tail behavior  similar to the covalently bonded state.  However, we find no clear indication of secondary peaks or features that have been associated with tunneling in a variety of experiments~\cite{kdp,supercool,reitersilica,proteinpdist}.  Additionally, we present the resultant mometum distributions of a particle in a one-dimensional double well potential that show that secondary features may indeed be observed when tunneling is ground-state dominated.

This article is organized as follows.  In Section \ref{sec:openpimd} we review the methodology of first principles open path integral molecular dynamics.  In Section \ref{sec:details} we discuss  the three high pressure ice systems under study and the simulation details, and then present the results of these computations in Section \ref{sec:hpresults}.  An open path integral molecular dynamics simulation of a simple one-dimensional model of a particle in a double-well potential is presented in Section \ref{sec:1d} and discussion and conclusions are given in Section \ref{sec:conc}.

\section{Open path integral Car-Parrinello molecular dynamics} \label{sec:openpimd}
The momentum distribution, $n(\bm{p})$, may be written as the Fourier transform of the density matrix in position space, $\rho(\bm{r},\bm{r}^\prime)$:
\begin{eqnarray}
n(\bm{p}) &=& \int \mathrm{d}\bm{r} \mathrm{d}\bm{r}^\prime \, e^{-\frac{\imath}{\hbar} \bm{p}\cdot(\bm{r}-\bm{r}^\prime)} \rho(\bm{r},\bm{r}^\prime) \label{eqn:pdist}
\end{eqnarray}
The path integral discretization of the density matrix maps the quantum system onto a set of $P$ replicas (``beads'')  that obey classical physics, thereby allowing one to utilize the machinery of computational classical statistical mechanics, namely Monte Carlo and molecular dynamics.  The discretized density matrix may be written as:
\begin{eqnarray}
\rho(\bm{r},\bm{r}^\prime) & = & \lim_{P \rightarrow \infty} \int\limits_{\substack{\bm{r}_1=\bm{r} \\ \bm{r}_{P+1}=\bm{r}^\prime}} \mathrm{d}r_2 \ldots \mathrm{d}r_P \, e^{-\beta U_{\mathrm{eff}}} \label{eqn:pi_dmat}\\
U_{\mathrm{eff}} & = & \sum_{i=1}^P \frac{mP}{2\hbar^2\beta^2} \left| \bm{r}_i - \bm{r}_{i+1} \right|^2 \nonumber \\
 && + \frac{V(\bm{r}_1) + V(\bm{r}_{p+1})}{2P} + \sum_{i=2}^P \frac{V(\bm{r}_i)}{P} \label{eqn:ueff}
\end{eqnarray}
This expression holds for a single particle but can be easily extended to multi-particle systems.

Typically, one is interested in computing equilibrium averages of position-dependent properties.  In this case, only diagonal elements of the density matrix are required and ``closed'' paths (for which $\bm{r}=\bm{r}^\prime$) must be sampled.  However, the Fourier transform shown in Equation \ref{eqn:pdist} requires off-diagonal components, and may be computed via ``open'' paths.  In this formalism, the momentum distribution is related to the Fourier transform of the end-to-end distance distribution of the open path.

In order to improve the efficiency of sampling, closed path integral molecular dynamics simulations may employ a coordinate transformation on $\{ \bm{r} \}$ that decouples the harmonic interactions in the first term of Equation \ref{eqn:ueff}.  One such approach is the staging transformation~\cite{cep_stage,sprik1,sprik2,pimd1}.  The staging transformation has been recently extended to open path integral simulation~\cite{morrone}.  First, the following identity is employed:
\begin{eqnarray}
\prod\limits_{i=1}^{P} && \hspace{-0.09in} e^{-\frac{mP}{2\hbar^2\beta} |\bm{r}_i - \bm{r}_{i+1}|^2} = \nonumber \\
&& \prod\limits_{i=2}^P \; e^{-\frac{m_i^\text{st} P}{2\hbar^2\beta}|\bm{r}_i - \bm{r}_i^*|^2 }  \times \, e^{-\frac{m}{2\hbar^2\beta} |\bm{r}_1 - \bm{r}_{P+1}|^2} \label{eqn:stagtform}
\end{eqnarray}
where:
\begin{eqnarray}
\bm{r}_i^* & = & \frac{ (i-1) \bm{r}_{i+1} + \bm{r}_1 }{i} \\
m_i^\text{st} & = & m \left(\frac{i}{i-1}\right) \, ,
\end{eqnarray}
and $m_i^\text{st}$ are known as the staging masses.  We can then transform to the following set of coordinates, $ \left\{ \bm{r}_1,\bm{u}_2 \ldots \bm{u}_{P+1} ,\ldots, \bm{r}_{P+1}\right\} $ with $\bm{u}_i =  \bm{r}_i - \bm{r}_i^*$ for $i=2,P$.  The two endpoints of the chain undergo a transformation into relative and center-of-mass coordinates:
\begin{eqnarray}
\bm{u}_{1} & = & \frac{\bm{r}_1 + \bm{r}_{P+1}}{2} \\
\bm{u}_{P+1}  & = & \bm{r}_1 - \bm{r}_{P+1}
\end{eqnarray}
The effective potential (Equation \ref{eqn:ueff}) may be written as:
\begin{eqnarray} 
U_\text{eff}^\text{ST}(\{ \bm{u} \}) &=& \frac{m}{2\hbar^2\beta^2} \bm{u}_{P+1}^2   + \sum\limits_{i=2}^{P}\frac{m_i^\text{st} P}{2\hbar^2\beta^2}\bm{u}_i^2 \nonumber \\*
&& + \, \frac{V(\bm{r}_1(\{\bm{u}\}))+V(\bm{r}_{P+1}(\{\bm{u}\}))}{2P} \nonumber \\
&&+ \, \sum\limits_{i=2}^{P} \frac{V(\bm{r}_i(\{\bm{u}\}))}{P} \label{eqn:openstueff} \\
\text{where:} & \{ \bm{u} \} = & \left \{ \bm{u_1},\bm{u}_2 \ldots \bm{u}_P,\bm{u_{P+1}}  \right \}
\end{eqnarray}
In order to generate molecular dynamics  trajectories the forces $\bm{f}_{\bm{u}} = -\frac{\partial V}{\partial {\bm{u}}}$ are computed as follows:
\begin{eqnarray}
\bm{f}_{\bm{u}_1} & = & \bm{f}_A + \bm{f}_B \\
\bm{f}_{\bm{u}_{P+1}} & = & \frac{\bm{f}_A - \bm{f}_B}{2} \\
\bm{f}_{\bm{u}_i} & = & \frac{1}{P}\bm{f}_{\bm{r}_i} + \left( \frac{i-2}{i-1} \right) \bm{f}_{\bm{u}_{i-1}} \; \; \; (i=2,P)  \label{midf}
\end{eqnarray}
where:
\begin{eqnarray}
\bm{f}_A &=& \frac{1}{2P}\bm{f}_{\bm{r}_{P+1}} + \frac{1}{P}\sum\limits_{i=2}^P \left(\frac{i-1}{P}\right) \bm{f}_{\bm{r}_i} \\
\bm{f}_B &=& \frac{1}{2P} \bm{f}_{\bm{r}_1} + \frac{1}{P}\sum\limits_{i=2}^P \left(\frac{P-i+1}{P}\right) \bm{f}_{\bm{r}_i} 
\end{eqnarray}

Staging open path integral molecular dynamics has been employed in conjunction with the Car-Parrinello~\cite{cp1} scheme in Reference \onlinecite{morrone2}.  This is a straightfowared extension of pre-existent  closed path integral Car-Parrinello molecular dynamics~\cite{marx1,marx2,cppimd}.  The corresponding extended Lagrangian is given by:
\begin{eqnarray}
\mathcal{L} &=& \sum\limits_{i=1}^{P+1} \Bigg\{ \frac{1}{2}m^\prime_i \left|\dot{\bm{u}}_i\right|^2 + \sum\limits_{s} \frac{\mu}{gP} \int \text{d} \bm{r} \left| \dot{\phi}_i^{(s)}(\bm{r}) \right|^2 \Bigg\} \nonumber \\
&& -\Bigg\{\frac{m}{2\hbar^2\beta^2} \bm{u}_{P+1} + \sum\limits_{i=2}^P \frac{m^{st}_i P}{2\hbar^2\beta^2}\left|\bm{u}_i\right|^2 \Bigg\}\nonumber \\
&& + \sum\limits_{i=1}^{P+1} \Bigg\{ - \frac{1}{gP} E\left[ \{\bm{r}\}_i, \{\phi\}_i \right] \nonumber \\
&& + \sum\limits_{s,t} \Lambda_{st} \left[\int \text{d}\bm{r} \phi_i^{*(s)}(\bm{r}) \phi_i^{(t)} (\bm{r}) - \delta_{st} \right] \Bigg\}  \\
\text{with:} && \nonumber \\
&& g = \begin{cases}
 2 & \text{if $s=1,P+1$},\\
 1 & \text{otherwise}
 \end{cases}
\label{eqn:stpicpl}
\end{eqnarray}
where $\mu$ is the fictitious electron mass, $E$ is the ground state potential energy, and the final term enforces the condition of orthonormality upon the orbitals. The masses, $m_i^\prime$, are associated with the velocities of the staging coordinates and are chosen to be a multiple of their corresponding staging masses, $m_i^\text{st}$.

The momentum distribution is a single-particle property. For multi-particle systems within the open path integral formalism, it may be exactly computed when one path is opened and all others are closed.  This requirement would lead to very inefficient sampling in bulk materials.  However, it has been shown that if the paths of multiple particles are ``opened'' and these paths are sufficiently far apart from each other,  the impact upon the resultant distribution is negligible~\cite{morrone}.  For a system of water molecules, one hydrogen per molecule is treated with an open path. All oxygen and all other hydrogen paths are closed. The end-to-end path distribution is then averaged over all open paths.

\section{Simulation Details} \label{sec:details}
Ice possesses a rich phase diagram.  At ambient pressure and below O$^\circ$C, water is most stable in a hexagonal crystal structure.  This is the form of ice for which the momentum distribution has been previously studied~\cite{burnham,morrone,burnham08,morrone2}.  However, under conditions of very high pressure, individual water molecules are arranged in interpenetrating cubic hydrogen bonded lattices.  This arrangement forms an effective body centered cubic (BCC) lattice structure.  In this study, we will concetrate on three phases, Ice VII, VIII, and X.  Ice VIII is proton ordered, exhibiting an anti-ferroelectric hydrogen bonding pattern.  In comparison, Ice VII is proton disordered.  Under higher pressures, the oxygen-oxygen distance reduces to the point where the proton's most stable position is equidistant between oxygen atoms and is located at the midpoint of the hydrogen bond axis.  This ``symmetric'' form of ice is known as Ice X~\cite{lee2,lee1}.

\begin{figure}[h]
\begin{center}
\includegraphics[scale=0.30]{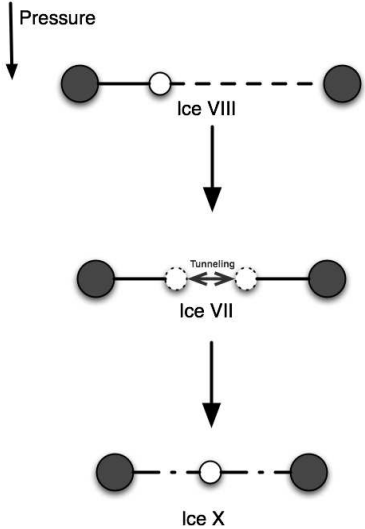}
\caption{A schematic of the atoms involved in a single hydrogen bond in the three high pressure ice phases presently under study. The gray circles represent oxygen atoms and the white circles represent hydrogen.  As the pressure upon the system increases the average oxygen-oxygen distance decreases, which has important consequences for the state of the proton.  This may be covalently bonded (Ice VIII), tunnel between wells (Ice VII) or lie in a symmetric state between the oxygen atoms (Ice X).}  \label{fig:cartoon}
\end{center}
\end{figure}

The work of Benoit and Marx~\cite{icecpmd,marxice2} has shown that by varying the lattice parameter (which changes the volume, and is equivalent to a change in pressure) of an Ice VIII cell one may, after a suitable equilibration period, generate Ice VII and Ice X.  In the case of Ice VII, the system will tunnel through the barrier along the hydrogen bond axis, thereby disrupting the proton ordering in the system.  At even smaller volumes, Ice X becomes thermodynamically favored.  A schematic is provided in Figure \ref{fig:cartoon} that illustrates these concepts.

Presently, we consider a $2\times 2\times 2$ BCC supercell containing 16 water molecules at three different lattice constants.  The lattice constants, as well as the corresponding molar volumes, pressures, and most probable oxygen-oxygen nearest neighbor distance are given in Table \ref{tab:hpparms}.   The approximate pressures are garnered from the equation of state given by Hemley \emph{et al.}~\cite{hemley}.

The first principles open path methodology is employed in order to generate the trajectories.  After an equilibration of 4 ps, each system is simulated for 75 ps, with the exception of System 2, which is sampled for 120 ps.  A timestep of 0.0725 fs is employed in all simulations.  Each system is sampled at 100K.  The temperature is controlled by means of massive Nose-Hoover chain thermostats~\cite{nose1,hoover,nhc1}.  Each path contains 32 replicas. The electronic states are evolved utilizing the Car-Parrinello methodology~\cite{cp1} with a fictitous mass of 340 atomic units.  The electronic structure is described via the Kohn-Sham formulation of Density Functional Theory~\cite{kohnsham} where exchange and correlation effects are treated by the  BLYP functional~\cite{becke,leeyangparr}.  The valence orbital are expanded in a plane wave basis set with a cutoff of 75 Rydberg.  Troullier-Martins norm-conserving pseudopotentials~\cite{tmpp} are utilized to model the valence effects of core electrons.  The dynamical masses associated with the staging coordinates are set to be a factor of 4 larger than the staging masses.

Despite the small number of water molecules in the simulation cell, there are 2048 electron states (32 replicas $\times$ 16 molecules $\times$ 4 states per molecule) present in the system.  This is a relatively large system by the standards of first principles simulation, and only state-of-the-art computational resources make possible the calculation of the relatively long trajectories and multiple systems reported in this study.  All computations are performed on IBM Blue Gene/L hardware with the {\sc cpmd} program~\cite{cpmd}, which has been optimized for this archetecture~\cite{curioni1,curioni2}.

\begin{table}[h] 
\begin{center}
\begin{tabular}{||c|c|c|c|c||}
\textbf{System} & \textbf{Lattice}      & \textbf{Molar}    & \textbf{Approx.} & \textbf{d$_{OO}^\text{mp}$} \\
\textbf{Number} & \textbf{Constant} &\textbf{Volume} & \textbf{Pressure}  &   \\ \hline
1 &2.67 $\text{\AA}$ & 5.74 cm$^3$/mol &  90 GPa & 2.31$\text{\AA}$ \\
2 & 2.84 $\text{\AA}$ & 6.90 cm$^3$/mol&  45 GPa & 2.45$\text{\AA}$ \\ 
3 & 2.94 $\text{\AA}$ & 7.62 cm$^3$/mol & 31 GPa &2.53$\text{\AA}$ 
\end{tabular}
\caption{Characteristic values that relay the size of each 16 molecule high pressure ice cell are given in the table above.  The pressure is approximated from the equation of state given by Hemley \emph{et al.}~\cite{hemley}  The value of d$_{OO}^\text{mp}$ is the most probable oxygen-oxygen distance between nearest neighbor, hydrogen bonded molecules.} \label{tab:hpparms}
\end{center}
\end{table}

\section{High Pressure Ice} \label{sec:hpresults}

The distributions in position and momentum space are computed in each system.  As noted in Section \ref{sec:openpimd}, open paths are utilized for the computation of the momentum distribution, and closed paths are appropriate for the position distribution.  Since our simulation contains both open and closed paths, we are able to use the closed paths for position distributions, and the open paths for the computation of the momentum distribution.  The nature of this system in position space has already been elucidated in previous studies~\cite{icecpmd,marxice2,benoit05}.  Here we repeat this work in order to explore the relation between the position and momentum space distributions.

\begin{figure}[h]
\begin{center}
\includegraphics[scale=0.35]{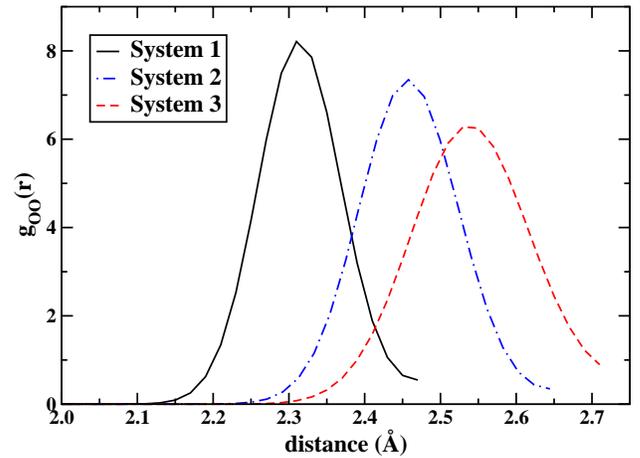}
\caption{(Color online) The first peak of the oxygen-oxygen radial distribution function in System 1 (solid curve), System 2 (dot-dashed curve) and System 3 (dashed curve). As one would expect, as the molar volume is decreased, the nearest neighbor oxygen-oxygen distance is as well.} \label{fig:goo}
\end{center}
\end{figure}

In Figure \ref{fig:goo}, the first peak of the oxygen-oxygen radial distribution function is shown.  Shortening of the oxygen-oxygen distance is apparent as the molar volume is decreased.    The position of the first peak of each distribution is reported in Table \ref{tab:hpparms}.  Although there is roughly two-tenths of an angstrom difference between oxygen-oxygen distances of Systems 1 and 3, this has a dramatic impact upon the nature of the proton that is confined on the potential energy surface.  It is this shortening that drives the phase transition between the forms of ice under study.~\cite{icecpmd,marxice2,benoit05}

\begin{figure}[h]
\begin{center}
\includegraphics[scale=0.35]{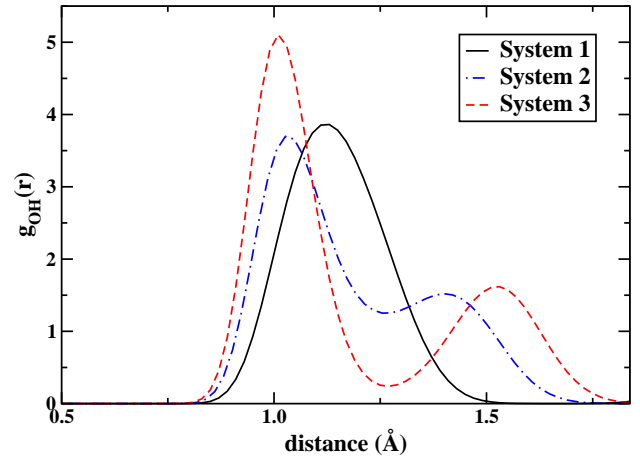}
\caption{(Color online) The oxygen-hydrogen radial distribution function in System 1 (solid curve), System 2 (dot-dashed curve) and System 3 (dashed curve).  Whereas in System 3 there is a distinction between covalent and hydrogen bonding distances, the two peaks have merged in System 1.  } \label{fig:goh}
\end{center}
\end{figure}

\begin{figure}[h]
\begin{center}
\includegraphics[scale=0.35]{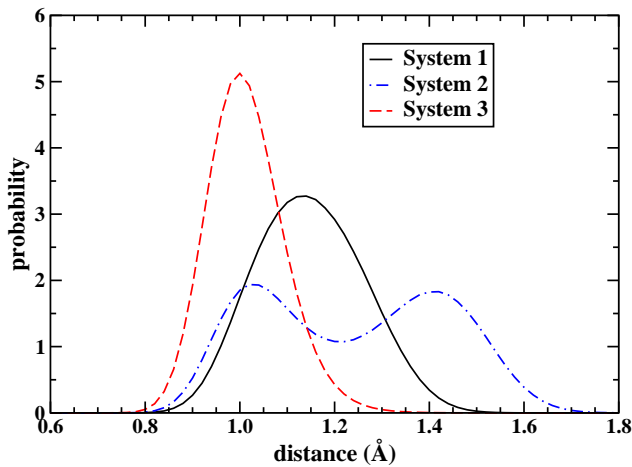}
\caption{(Color online) The distance distribution of the proton along the oxygen-oxygen direction in System 1 (solid curve), System 2 (dot-dashed curve) and System 3 (dashed curve).  This direction is analogous to the hydrogen bonding axis.   One may note that the distribution of System 2 is delocalized across two wells.} \label{fig:pz}.
\end{center}
\end{figure}

The position space distribution of the proton along the oxygen-oxygen hydrogen bond axis is illustrated by the oxygen-hydrogen radial distribution functions (Figure \ref{fig:goh}) and the probability distribution of the proton position along the hydrogen bond axis (Figure \ref{fig:pz}).  It can be seen that the proton in System 3 remains covalently bonded to its oxygen, although the covalent bond distribution is broader than in typical water phases.  This system retains the Ice VIII structure.  It can be seen in Figure \ref{fig:goh} that the covalent bond and hydrogen bond peaks of the radial distribution function of System 1 merge.  This broad single peak located at the midpoint between the two oxygen atoms is indicative of a symmetric hydrogen bond as found in the Ice X phase. 

Evidence of quantum tunneling can be seen in System 2.  The bimodal nature of the proton distribution in Figure \ref{fig:pz}, as well as the fact that one peak is near the covalently bonded peak of System 3 indicates that there are tunneling events from one well to another along the hydrogen bond axis.  It was shown in the work of Benoit and Marx~\cite{icecpmd,marxice2} that classical protons at this molar volume and temperature do not cross the barrier and remain trapped in a single well.  This calculation showed that thermal hopping over the barrier is disfavored and quantum tunneling dominates.  As noted in Section \ref{sec:details}, the tunneling disrupts the anti-ferroelectric ordering and engenders the formation of Ice VII.   We note that the bimodal distribution in Figure \ref{fig:pz} is not perfectly symmetric.  This may be caused by insufficient sampling or asymmetries that arise from correlated proton motions.

We note that the present distributions are somewhat more delocalized when compared with the work of Benoit and Marx~\cite{icecpmd,marxice2}.  This is likely a result of the use of a larger number of replicas in the present computation.  However there are many other differences in the details of the simulation that may impact this result, including trajectory length and the choice of exchange-correlation functional.  Overall however, the description of the proton in position space along the hydrogen bond axis is in good agreement with this and later work.~\cite{benoit05}

\begin{figure}[h]
\begin{center}
\includegraphics[scale=0.35]{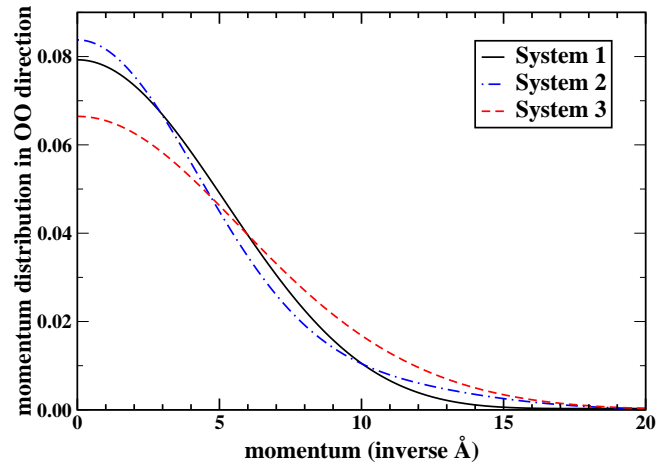}
\caption{(Color online) The proton momentum distribution in the oxygen-oxygen (OO) direction in System 1 (solid curve), System 2 (dot-dashed curve) and System 3 (dashed curve).  It is in this orientation that the distinctions between phases occur.  } \label{fig:pdistz}
\end{center}
\end{figure}

\begin{figure}[h]
\begin{center}
\includegraphics[scale=0.35]{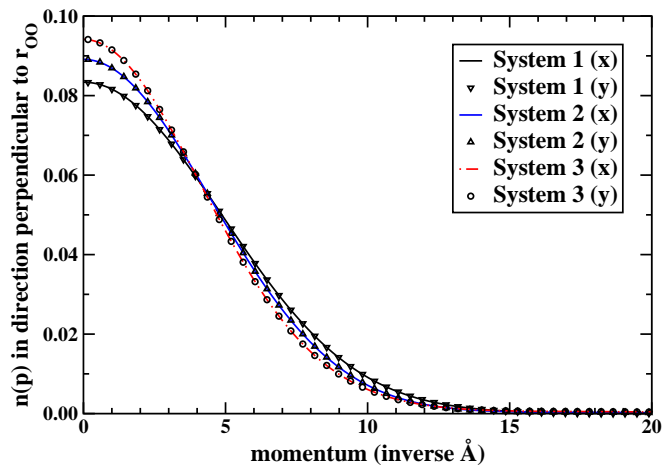}
\caption{(Color online) The proton momentum distribution perpendicular to the oxygen-oxygen direction (denoted ``x'') in System 1 (solid curve), System 2 (dot-dashed curve) and System 3 (dashed curve).  Also plotted are the proton momentum distributions in the mutually orthogonal direction (denoted ``y'') in System 1 (triangles pointing downward), System 2 (triangles pointing upward) and System 3 (circles). The differences in widths of these curves indicate the relative pressure upon each system.} \label{fig:pdistxy}
\end{center}
\end{figure}

The momentum distribution is plotted along the oxygen-oxygen axis (Figure \ref{fig:pdistz}), as well along the two corresponding perpendicular axes (Figure \ref{fig:pdistxy}). These are effective one-dimensional plots that are computed via the Fourier transform of the path end-to-end distance distribution along these directions.  In Figure \ref{fig:pdistxy}, one can view a trend that the momentum distributions in the directions perpendicular to the hydrogen bond broaden with decreasing system molar volume.  This is consistent with the uncertainty principle given that as the protons become more confined in position space, the corresponding momentum distributions have a greater variance.   Aside from this difference, there is little distinction between the systems under study in these directions when compared to the momentum distribution projected onto the hydrogen bonding axis (see Figure \ref{fig:pdistz}).  This is a logical conclusion as the large qualitative differences in position space occur in the hydrogen bond direction (see Figure \ref{fig:pz}), as shown presently and in previous work on high pressure ice~\cite{icecpmd,marxice2,benoit05}.

One may also note in Figure \ref{fig:pdistxy} that the distributions are similar along the two directions perpendicular to the hydrogen bond axis.  This chemically intuitive result is in agreement with a previous study of the ``shape'' of the proton high pressure ice phases~\cite{benoit05}, where it was found that the position space distribution in the perpendicular directions were of similar extent. In addition, the proton's variance in the perpedicular directions was shown to decrease with increasing pressure~\cite{benoit05}, thereby providing complementary information to the momentum space picture discussed above.

The position space distributions show that System 1 contains symmetric hydrogen bonds, System 2 exhibits a bimodal proton distribution and in System 3, the protons are covalently bonded.  In Figure \ref{fig:pdistz}, we present the momentum distributions in the hydrogen bonding direction.  The covalently bonded System 3 possesses the narrowest position distribution (see Figure \ref{fig:pz}) and therefore the correspondingly broadest momentum distribution.  The high-momentum tail of this distribution is dominated by the OH stretch (see Section \ref{sec:intro}).  In the symmetric hydrogen bonded case (System 1), the more delocalized bond yields a narrower momentum distribution with a shortened tail.  This signature in proton momentum distributions corresponds to a red-shift of the OH stretching frequency in stronger hydrogen bonded environments. This has been observed previously in the experimental~\cite{reiter} and simulation momentum distribution of liquid water and hexagonal ice~\cite{morrone2} .  The symmetric hydrogen bond may be considered the ``strongest'' class of hydrogen bonding.  Such an interpretation is bourne out by experiments on symmetric hydrogen bonds observed in water confined in nanotubes~\cite{reiternano} and Rb$_3$H(SO$_4$)$_2$~\cite{homouz} that exhibit greatly narrowed momentum distributions with shortened tails.

The shape of the proton momentum distribution in the tunneling direction in System 2 lends to a more detailed description.  It appears to have an anomalous shape when compared to the other distributions.  Namely, it is narrow at low momentum, yet its tail behavior is similar to that of the covalently bonded System 3.  This tail behavior is likely engendered by the localization in the covalently bonded well that is a component of the tunneling system. Therefore the highest frequency components of the system are similar to those exhibited in System 3.  The narrowness exhibited in the low-momentum region is related to the overall delocalized nature of the proton.  The tunneling momentum distribution will be further investigated in Section \ref{sec:1d}.  

\begin{figure}[h]
\begin{center}
\includegraphics[scale=0.37]{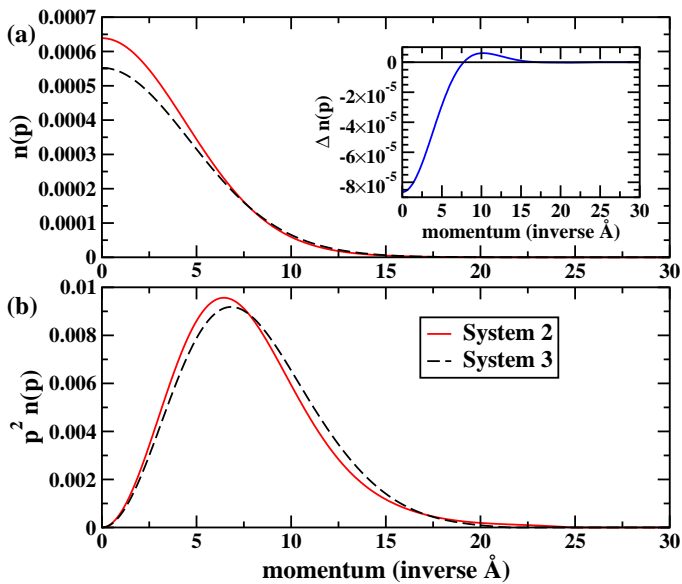}
\caption{(Color online) The spherically averaged momentum distributions and radial momentum distributions of Systems 2 (solid curve) and 3 (dashed curve) are plotted in panel (a) and panel (b), respectively.  Each curve is normalized such that area under $4 \pi p^2 n(p)$ is equal to one.  The difference between the plotted momentum distributions is depicted in the inset of panel (a).} \label{fig:spher}
\end{center}
\end{figure}

In Figure \ref{fig:spher}, the spherically averaged momentum distribution $n(p)$, and the radial momentum distribution, $p^2 n(p)$ are depicted for Systems 2 and 3.  Note that the difference between the distributions is dominated by distinctions present in the hydrogen-bonding direction, although this is somewhat muted by the contributions from the perpendicular orientations.

\begin{figure}[h]
\begin{center}
\includegraphics[scale=0.40]{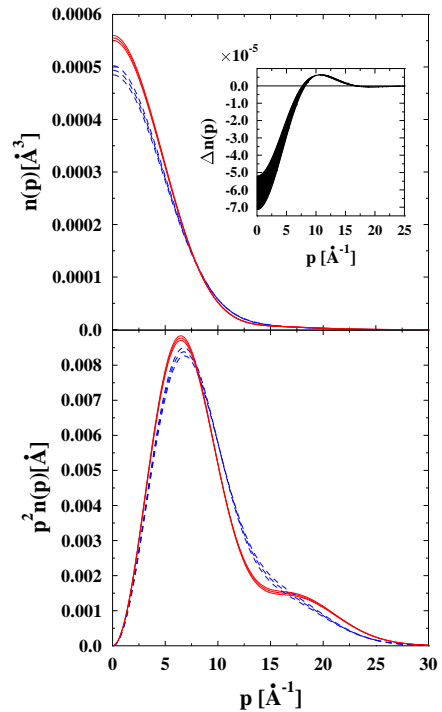}
\caption{(Color online) The experimental proton momentum distribution of hydrated lysozyme.  These results were reported in Figure 3 of Senesi \emph{et al.}~\cite{proteinpdist} The spherically averaged and radial momentum distributions of a proton in a single well (dashed curve) and double well potential (solid curve) are plotted in the upper and lower panel, respectively.  The normalization matches that of Figure \ref{fig:spher}.  The curves above and below the distributions delineate the error in the measurement of $n(p)$~\cite{noreland}.  The difference between the plotted momentum distributions is depicted in the inset of the upper panel. Figure courtesy of R. Senesi. } \label{fig:exp}
\end{center}
\end{figure}

The change in the spherically averaged momentum distribution of Figure \ref{fig:spher} reflects the transition in the state of the proton from System 2 (tunneling) to System 3 (covalently bonded).  Interestingly, this change bears similarity to that observed in the measured spherical momentum distribution of a proton in the hydration shell of the globular protein lysozyme~\cite{proteinpdist} at different temperatures.  Although these are different systems, the momentum distribution is mostly dependent upon the local environment of the proton.  Therefore it is likely that qualitative features are common among the set of proton tunneling systems.  To facilitate the comparison, we report the experimental distributions in Figure \ref{fig:exp}.  The similarity between theory and experiment supports the interpretation that the observed change in the momentum distribution signals the onset of tunneling behavior.  In our simulation, this is caused by a change of pressure whereas in the experiment, it results from a change in temperature.  In both cases (see the insets of Figures \ref{fig:spher} and \ref{fig:exp}) the momentum distribution of a proton in a single well differs from that in a double well potential via a narrowed distribution in the low-momentum region. In addition, the experimental tunneling distribution displays a distinct feature in the tail that is only visible when the distribution is multiplied by a factor of $p^2$.  This effect cannot be detected in our simulation.  This may reflect genuine differences in the state of the proton in the experiment and in the present simulation.  However, one should note that such small differences in the tail of the momentum distribution are beyond the precision of the current computation. 

In addition to the work of Senesi \emph{et al.}~\cite{proteinpdist}, experimental studies have reported secondary features in the momentum distribution~\cite{kdp} or radial momentum distribution~\cite{supercool,reitersilica} in a variety of other systems, where they have also been interpreted as a signature of tunneling. The presence of secondary features in tunneling systems will be studied further in the next section.

Experimental results such as those shown in Figure \ref{fig:exp}  manifest excess kinetic energy in comparison to liquid water and hexagonal ice~\cite{proteinpdist,reitersilica,supercool}.  This is particularly expressed in broadened tails of the momentum distribution.  In some systems, this phenomenon has been identified with tunneling modes~\cite{reitersilica,supercool}.  In contrast, excess kinetic energy is not apparent in the simulation results that are depicted in Figure \ref{fig:spher}. Instead, the tail of the distribution in Systems 2 and 3 remains dominated by covalent bond stretching frequencies like those exhibited in the liquid and hexagonal crystal phases~\cite{reiter,newexp,morrone2}.  However, we note that from the simulation perspective, the kinetic energy may be most accurately computed from \emph{closed} path integral simulations~\cite{virial}, as the tail of the momentum distribution is difficult to obtain to high precision from the present methodology.

\section{A simple model for tunneling} \label{sec:1d}

In order to further investigate the tunneling behavior of the proton, we present an analysis of a particle in a one-dimensional potential.  This degree of freedom corresponds to the displacement of the proton along the hydrogen bond axis as plotted in Figures \ref{fig:pz} and \ref{fig:pdistz}.  Such models are often employed in the literature~\cite{hbondmodels}, although we do not claim all the complexities of the high pressure ice system can be reduced to an effective one-dimensional form.   Instead, it is simply a tool to easily investigate the equilibrium distributions of position and momentum in the tunneling regime.  Therefore, we limit our discussion to an analysis of these properties and do not consider tunneling kinetics or dynamics.  These issues have been extensively studied in the literature.~\cite{weiss87,gillan87,voth89,topaler,jang99,jang2000,ramirez04}

\begin{figure}[h]
\begin{center}
\includegraphics[scale=0.60]{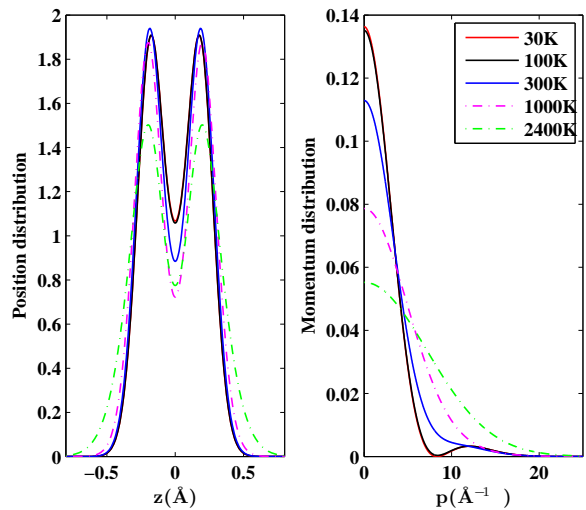}
\caption{(Color online) The position (left panel) and momentum (right panel) distributions of the double well model at several temperatures. One may note that there are larger qualitative differences in the momentum space picture than the position space picture with the increase in temperature.  Note the disappearance of secondary features in the momentum distribution which is evident at $T=300$K. } \label{fig:1d} 
\end{center}
\end{figure}

We utilize a potential of the following form:
\begin{eqnarray}
V(z) = \frac{1}{2} m \omega^2 z^2 + A e^{-m\xi z^2} \label{eqn:1d}
\end{eqnarray}
with $m=1836$, $\omega=0.005$, $A=0.012$, and $\xi=0.0087$.  The parameter $\omega$ characterizes the confinement of particle in the absence of the barrier, and $A$ and $\xi$ define the height and width of the potential barrier, respectively. All parameters are reported in atomic units and were chosen to yield a similar distribution to System 2 in position space (see Figures \ref{fig:pz} and \ref{fig:1d}).  Our potential choice in Equation \ref{eqn:1d} is by no means unique, as other model potentials may be employed~\cite{hbondmodels}.  

\begin{figure}[h]
\begin{center}
  \includegraphics[scale=0.80]{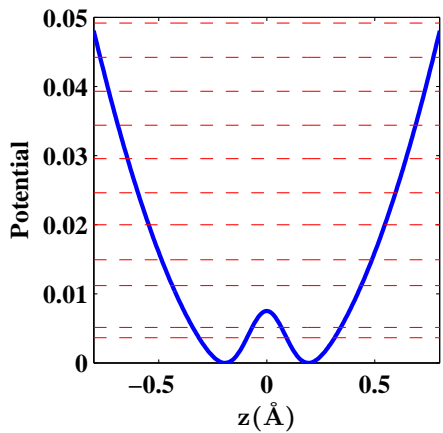}
\end{center}
\caption{(Color online) The model double well potential (solid curve) is plotted alongside the first eleven energy levels of the spectra (dashed lines).  Note that the ground and first excited state lie below the barrier. }
\label{fig:potential}
\end{figure}

Given that the system is a single particle in a one-dimensional potential, we may directly diagonalize the system Hamiltonian in order to obtain the spectra.  There are a large variety of schemes available that numerically solve the one-dimensional Schr\"odinger equation~\cite{PryceAndrew1993}.  We have utilized {\sc sledge}~\cite{PruessFulton1993} for the spectral computations presented here.  The shape of the potential barrier is plotted against the lowest eleven energy levels in Figure \ref{fig:potential}.  It can be seen that the ground state and the first excited state lie below the barrier.  Higher energy levels are nearly equally spaced as the harmonic term of Equation \ref{eqn:1d} begins to dominate.

The distributions in position and momentum space may be computed as a summation over the density matrix:
\begin{eqnarray}
n(x) = \frac{1}{Z}\sum_{i=0}^{\infty} e^{-\beta E_i} |\varphi_{i}(x)|^2 \\
n(p) = \frac{1}{Z} \sum_{i=0}^{\infty} e^{-\beta E_i} |\varphi_{i}(p)|^2
\end{eqnarray}
where the partition function is $Z=\sum_{i=0}^{\infty}e^{-\beta E_i}$.  This infinite summation is truncated when $ e^{-\beta (E_i - E_1)} < 0.001$.  The populations of the lowest two energy states as well as the population of states above the barrier are given in Table \ref{tab:pop}.  The position and momentum distributions at various temperatures are depicted in Figure \ref{fig:1d}.  

Upon inspection of the position distributions, one notes that the bimodal shape indicates there are barrier crossings at all temperatures.  At low temperatures, only states below the barrier are populated, and quantum tunneling is responsible for the traversing of the barrier.  As the temperature increases, thermally activated crossings begin to contribute and there is a crossover from quantum to classical behavior~\cite{gillan87}.  This crossover is not apparently detectable from the position distribution unless further simulations are performed on classical particles~\cite{icecpmd} or isotope effects are studied~\cite{radu02}.

The momentum distribution provides complementary information to the position distribution and about the impact of quantum tunneling upon the system.  In Figure \ref{fig:1d}, one finds that similar position space distributions may lead to rather different momentum distributions. A node is present in the momentum distribution at very low temperatures where the ground state dominates~\cite{nodenote}.  These distributions are qualitatively similar to those obtained for KDP in neutron Compton scattering experiments~\cite{kdp}, as well as those garnered from ground state models that are described in the literature~\cite{reiter,kdp,reitersilica}.  The node disappears at intermediate temperatures where a non-gaussian momentum distribution with an extended tail is present.  This is the general shape of the momentum distribution of System 2 (see Figure \ref{fig:pdistz}).  At higher temperatures the distribution becomes more gaussian-shaped in the regime of the quantum/classical crossover.  

In addition, we note that the presence of nodes indicates a barrier in the underlying potential energy surface but not necessarily quantum tunneling of the ground state. A perturbation analysis predicts a node in the ground state for any barrier of finite height, even in the case where the barrier is ``washed out'' by zero-point-motion and a unimodal position distribution is present  (see Appendix \ref{sec:perturb}).   However, for systems that exhibit bimodal position distributions, a node in the momentum distribution is a signature of ground (``coherent'') state tunneling. Mixed (``incoherent'') state tunneling is indicated by a momentum distribution of shape similar to that in the model at $T=300K$ (Figure \ref{fig:1d}) or in System 2 (Figure \ref{fig:pdistz}).

\begin{table}[h]
\centering
\begin{tabular}{c||c|c|c|c|c|c|c}
$T$ & $30 \text{K}$ & $100\text{K}$ &  $300\text{K}$ & $1000\text{K}$ & $2400\text{K}$ \\
\hline
$P_0$ &   100\% &  99.1\% &  82.8\%  &  57.1\% &    37.5\% \\
$P_1$ &  0.00\% & 0.881\% &  17.2\% &  35.6\% &   30.8\%\\
$P_{2+}$ & 0.00\% & 0.00\% & 0.00\% &  7.29\% &   31.7\%\\
\end{tabular}
\caption{This table depicts the populations of the ground state, first excited state, and all other levels as a function of temperature for a single particle in the potential given in Equation \ref{eqn:1d}.}
\label{tab:pop}
\end{table}

\begin{figure}[h]
\begin{center}
\includegraphics[scale=0.30]{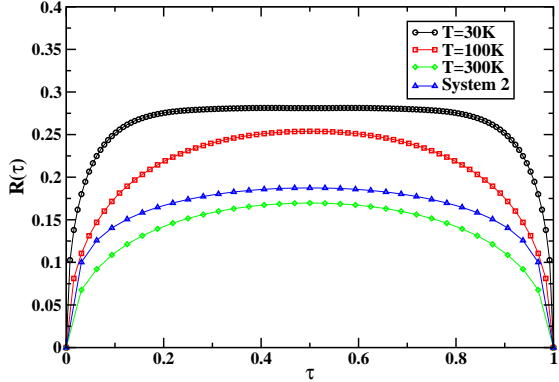}
\caption{(Color online) The root mean square imaginary time correlation function of the model system at T=30K (dotted curve with circles), T=100K (dotted curve with squares), T=300K (dotted curve with diamonds), and System 2 of the high pressure ice calculation (dotted curve with triangles).  The imaginary time is reported in units of $\beta \hbar$.} \label{fig:rms}
\end{center}
\end{figure}

To further investigate the temperature dependence of the model system and its relation to the high pressure ice calculation, open and closed path integral molecular dynamics simulations were undertaken at T=30K, T=100K, and T=300K.  Note that at each of these temperatures, there is virtually no population of energy levels above the barrier and tunneling effects dominate.  The position and momentum distributions garnered from these computations are in good agreement with those presented in Figure \ref{fig:1d}.  The path may be characterized by the root mean square imaginary time correlation function~\cite{chandler94,icecpmd,tuckerman01}, $\mathcal{R}(\tau)$.
\begin{eqnarray}
\mathcal{R}^2(\tau) &=& \left< \left[ z(\tau)-z(0)\right]^2\right> \;\;\;\; 0\leq \tau \leq \beta \hbar
\end{eqnarray}
The time independence (``flatness'') of the root mean square imaginary time correlation function near the midpoint is related to the degree of ground state dominance~\cite{chandler94}.  In Figure \ref{fig:rms} we plot this function for the three simulated temperatures and System 2.  The T=30K curve is the flattest, indicating that the system is in the ground state, whereas the others show some time dependence that is suggestive of the presence of some excited state weighting in the density matrix.

\begin{figure}[h]
\begin{center}
\includegraphics[scale=0.32]{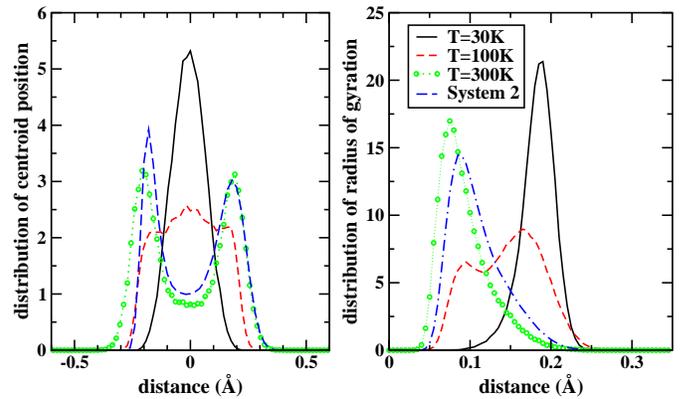}
\caption{(Color online) The distribution of the centroid of the paths (left panel) and the distribution of the radius of gyration (right panel) of the double well model at T=30K (solid curve), T=100K (dashed curve) and T=300K (dotted curve with circles) plotted against that of System 2 in the hydrogen bonding direction (dot-dashed curve).} \label{fig:rgcent}
\end{center}
\end{figure}

In addition to indicating ground state dominance, the height of root mean square imaginary time correlation functions is related to the degree of path localization.  In order to further characterize this property,  the distribution of the centroid and the radius of gyration~\cite{lee99,hayashi} of the proton paths is plotted in Figure \ref{fig:rgcent}.  The centroid is defined as the center of mass of the path:
\begin{eqnarray}
z_c &=& \frac{1}{P}\sum\limits_{i=1}^P z_i
\end{eqnarray}
and the radius of gyration is defined as:
\begin{eqnarray}
z_\text{gyr} &=& \frac{1}{P}\sum\limits_{i=1}^P \left|z_i - z_c\right|^2
\end{eqnarray}
In this way the average and variance of the path may be characterized.  


In the plots of the radius of gyration and the centroid, we see qualitatively distinct behavior in the one-dimensional model at each temperature.  When the system is in the ground state (T=30K), the broad paths are centered about the midpoint of the position distribution and delocalized across the two wells.  This finding is in agreement with Feynman-Kleinert perturbation analysis of a double well potential that display a minimum in the centroid potential of mean force at the barrier at zero temperature~\cite{kleinert}.  In the high temperature case, the spread of the path has greatly narrowed, and its centroid distribution is now bimodal, centered about the two wells of attraction.  Therefore, the tunneling appears to be occurring by two different ``mechanisms'' in the ground (30K) and mixed states (300K), a single delocalized species in the former and ``path hopping'' in the latter.  At 100K, the system is in an intermediate state between these two situations, as can be starkly seen in the bimodal form of its radius of gyration distribution (see Figure \ref{fig:rgcent}).  The path hopping at higher temperatures is made favorable by the stiffened harmonic interactions between beads at higher temperatures that force the localization of the path.  In the high temperature (classical) limit, one would expect the path to collapse to a point and thermally hop between wells.  The behavior at 300K seems to be approaching this limit, although as remarked earlier, quantum tunneling still dominates.  Delocalized paths centered at the top of the barrier and localized paths in the potential wells of a double well potential were characterized in the work of Tuckerman and Marx~\cite{tuckerman01}, and this is in good agreement with the picture presented here.

The centroid and radius of gyration distributions of System 2 along the hydrogen bonded axis is plotted in Figure \ref{fig:rgcent}.  One may see that although this simulation was performed at 100K, the results are qualitatively similar to those at 300K.  Both System 2 and the 300K model momentum distributions lack secondary features and have similar shapes (see Figures \ref{fig:pdistz} and \ref{fig:1d}).  The similarity in position and momentum space with the model at 300K suggests that System 2 is in a mixed state tunneling regime.  We remark that although the model potential of Equation \ref{eqn:1d} was tuned to mimic System 2, there are still effects that cannot be captured by the model, such as the proton correlations that arise from the enforcement of ice rule.  Despite these limitations, the model potential provides a very useful qualitative picture of the temperature dependence of tunneling phenomena.  The ``crossover'' from ground state to mixed state tunneling, as indicated by the population of the energy levels, is apparent in the shape of the momentum distribution, as well as the spread and centroid of the particle's path integral representation.

Finally, we remark that our one-dimensional model is characterized by a potential that does not depend on temperature.  Within this model, tunneling features in the momentum distribution become more pronouced at lower temperatures.  By contrast, experiments on systems such as KDP~\cite{kdp} and the hydration shell of lysozyme~\cite{proteinpdist} have reported secondary tunneling features that only exhibit themselves \emph{above} a critical temperature.  This behavior must be facilitated by a structural transition above the critical temperature in which the potential experienced by the proton, and particularly the tunneling barrier, changes significantly.  Furthermore, we note that the experimental interpretations rest in part on a ground state model~\cite{kdp,reitersilica}, in agreement with the present finding that nodes in the momentum distribution are associated with ground state tunneling.

\section{Conclusions} \label{sec:conc}
In this work we have presented an investigation of the position and momentum space distributions of the proton in tunneling and symmetric hydrogen bonded systems. Novel first principles open path integral molecular dynamics algorithms were utilized in order to compute the momentum distributions. Three phases of high pressure ice were studied at 100K.  Each phase typifies a qualitatively different state of the proton, covalently bonded (Ice VIII), tunneling (Ice VII), and equally shared between nearest-neighbor oxygens (Ice X).

In Ice X, symmetric hydrogen bonds are present.  This phase is characterized by a broadened position space when compared to the covalently bonded Ice VIII distribution.  In accordance with the uncertainty relation between position and momentum, the computed momentum distribution is narrowed in the direction of the hydrogen bond axis.  This is also in agreement with the experimental signatures of symmetric hydrogen bonding that have been observed in other systems~\cite{reiternano,homouz}.  Furthermore, this phenomenon is an extreme case of the narrowing of the proton momentum distribution that accompanies the red-shift in the oxygen-hydrogen stretching frequency of hexagonal ice relative to liquid water~\cite{reiter,morrone2}.

The resultant proton momentum distribution of the tunneling system (Ice VII) also shows some qualitative correspondence to what has been observed experimentally.  The tunneling distribution is narrowed when compared to the covalently bonded distribution, in accordance with the relatively delocalized bimodal distribution that is a signature of tunneling in position space.  The tail behavior is similar to that of the covalently bonded distribution, indicating that the high frequency associated with the covalent bond is still present in the system.  However, we did not observe tunneling signatures such as nodes or secondary features in the momentum distribution that have been dectected in experiment.  As noted in Section \ref{sec:hpresults}, such features may be rather subtle~\cite{proteinpdist} and require sensitivity in the tail of the momentum distribution where the intensity is very small.  Such precision is beyond the scope of the currently utilized methodology, and addressing this issue is an important goal of future algorithmic development.   Furthermore, such features intimately depend upon the details of the potential experienced by the proton.  Direct comparison of theory and experiment for a system in which these signatures are unambiguously detected should therefore provide an extremely accurate test of the current description of hydrogen bonding.

To better understand how tunneling affects the momentum distribution, we performed an open path integral simulation of a particle confined in a one-dimensional double well potential. The position and momentum distributions may be directly computed in this system from the spectrum of eigenvalues and eigenvectors.  We found that a node is present in the momentum distribution at low temperature, which is indicative of ground (``coherent'') state tunneling.  In this case, the form of the distribution is in qualitative agreement with experimental studies of potassium dihydrogen phosphate~\cite{kdp}.  The Ice VII momentum distribution bears a resemblance to the results of the one-dimensional model at higher temperaure (300K), where the node is ``washed out.''  A clue as to why is revealed upon inspection of the centroid and radius of gyration distributions of the path, which show that the path tends to be localized in the wells rather than delocalized across the domain, as path integral simulations of the low-temperature (node-containing) one-dimensional systems.  The relative localization in position space lends itself to a broader distribution that is somewhat closer to that of the covalently bonded system.  

Whereas in a one-dimensional double well potential a secondary feature is always present when coherent tunneling dominates, in a three dimensional system such features may be washed out in the spherically averaged distribution.  This suggests that these characteristics may be more easily detected in crystalline systems where the momentum distribution can be measured along high symmetry directions.  On the other hand, the momentum distribution is very sensitive to the spatial environment experienced by the protons and simple one-dimensional models, while useful for identifying tunneling features, may be too crude for realistic predictions.  Indeed, the qualitative differences in the momentum distribution of the model and Ice VII at 100K, while certainly reflecting a crude parameterization of the model, may also indicate that important features of the potential energy surface of the high pressure ice system are not reducible to a single particle, one-dimensional form.

Finally, as noted in Section \ref{sec:intro} and discussed in Section \ref{sec:hpresults}, some experiments report excess kinetic energy as compared to ambient liquid water~\cite{reitersilica,supercool,proteinpdist}.  In this work, such behavior is not apparent.  All calculated tunneling distributions possess tail behavior that is similar to the covalently bonded Ice VIII (System 3).  The highest frequency in the system remains associated with the oxygen-hydrogen stretch, and no higher frequency modes act in the tunneling direction.  This result should be contrasted with the interpretation of recent experimental data suggesting that excess kinetic energy is associated with tunneling modes~\cite{reiter,reitersilica,supercool}.  Further investigations are necessary in order to unravel these findings.

\begin{acknowledgments}
We would like to acknowledge C. Andreani and R. Senesi for useful discussions and to thank R.S. for  providing us with Figure \ref{fig:exp}.  J.A.M. acknowledges the Fannie and John Hertz Foundation for its support of his graduate work.  Partial support for this work was provided by the DOE under grant DE-FG02-05ER46201 and by the NSF-MRSEC program through the Princeton Center for Complex Materials (PCCM), Grant DMR 0213706.  In addition, we would like to acknowledge Princeton University and IBM for the use of their computational resources.

This article has been submitted to the Journal of Chemical Physics.  After it is published, it will be found at http://jcp.aip.org.

\end{acknowledgments}

\begin{appendix}
\section{Perturbation analysis of the ground state momentum distribution}
\label{sec:perturb}

In this Appendix, we utilize a perturbation analysis in order to show that the ground
state momentum distribution always possesses extra nodes when a potential barrier exists. This is equivalent to showing that
the Fourier transform of the ground state has extra nodes.  

\begin{figure}[h]
\begin{center}
  \includegraphics[scale=0.6]{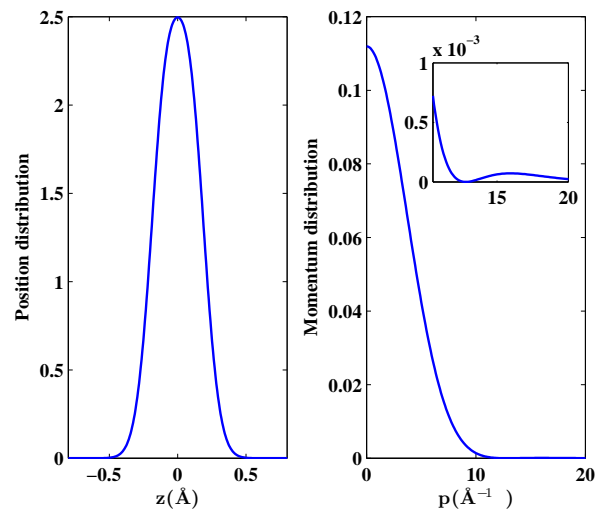}
\end{center}
\caption{(Color online) Ground state position (left panel) and momentum (right panel) distribution for the low barrier double well potential.  The inset depicts that the secondary feature persists in the momentum distribution even when zero-point motion washes out the barrier in position space.}
\label{fig:lowb}
\end{figure}

For example, this may be seen in Figure \ref{fig:lowb} where the momentum distribution of the model potential (Equation \ref{eqn:1d}) with $A=0.004$ and all other parameters unchanged from Section \ref{sec:1d} is plotted.  For these parameters, the barrier height is 300K and the ground state energy is 585K.  Therefore, zero-point motion washes out the barrier and the position distribution is unimodal.  However, a node is still detectable in the momentum distribution, though the secondary feature is diminished with respect to larger values of $A$.

Starting from a harmonic oscillator
\begin{equation}
V = \frac{1}{2}m\omega^2 z^2,
\label{}
\end{equation}
the eigenvalues are $E_n=(n+\frac12)\omega$, and corresponding 
eigenstates are denoted by $\ket{n}$. 
The ground state wavefunction in position space is:
\begin{equation}
\braket{z}{0} = \left(\frac{m\omega}{\pi}\right)^{1/4} \exp\left( -m\omega
z^2 / 2\right) H_0(\sqrt{m\omega}z),
\label{}
\end{equation}
The functions $H_n$ are Hermite polynomials.
The Fourier transform of the ground state is:
\begin{equation}
\braket{p}{0} = \left(\frac{m\omega}{\pi}\right)^{1/4}
\sqrt{\frac{1}{m\omega}}e^{-\frac{p^2}{2m\omega}}.  \label{}
\end{equation}
We are interested in the Fourier transform of the ground state 
in the presence of a potential perturbation:
\begin{equation}
\Delta V = A e^{-m\xi z^2}.
\label{}
\end{equation}
For simplicity, $m,\omega,\xi$ are chosen as before, and only $A$ is varied.

If $\abs{A}$ is small, the change in wavefunction can be captured by the
first order perturbation. Denote by $\ket{\varphi}$ the new ground
state, we have
\begin{equation}
\braket{p}{\varphi} = \braket{p}{0} + \sum_{i \text{ even and } i> 0} 
\frac{\braket{p}{i}\averageop{i}{\Delta V}{0}}{E_0 - E_i}.
\label{eqn:firstorder}
\end{equation}
Note that since the external potential is an even function all the odd terms in the sum vanish.  
We note that the infinite summation in Equation \eqref{eqn:firstorder} cannot be reduced to a simple form and is therefore difficult to use for further analysis. Despite the slow decay of this series, numerical investigation shows that qualitative information is already captured by including only the first two terms ($\ket{2}$ and $\ket{4}$) in the first order expansion.  In particular, the existence of the node in the presence of an arbitrary small potential barrier is already discernible within this approximation.

One finds:
\begin{equation}
\begin{split}
  \averageop{2}{\Delta V}{0} =&  
  \sqrt{\frac{1}{2^2 2!}} \left(\frac{m\omega}{\pi}\right)^{1/2}
  \int dz e^{-m\omega z^2}\\
  &H_2(\sqrt{m\omega} z)
  H_0(\sqrt{m\omega}z) A e^{-m\xi z^2}\\
  =& A \frac{1}{\sqrt{2}} \left(\frac{1}{\alpha^{3/2}} -
  \frac{1}{\alpha^{1/2}}\right)
\end{split}
\label{}
\end{equation}
Here $\alpha = 1+ \frac{\xi}{\omega}$.

Similarly:
\begin{equation}
  \averageop{4}{\Delta V}{0} = 
  A \frac{\sqrt{6}}{4} \left(\frac{1}{\alpha^{5/2}} -
  \frac{2}{\alpha^{3/2}}+\frac{1}{\alpha^{1/2}}\right).
\label{}
\end{equation}

Also we have:
\begin{equation}
\braket{p}{2} = 
\left(\frac{m\omega}{\pi}\right)^{1/4} 
\frac{1}{\sqrt{2}(m\omega)^{3/2}} e^{-\frac{p^2}{2m\omega}}
(m\omega-2p^2)
\label{}
\end{equation}

\begin{equation}
 \begin{split}
   \braket{p}{4} = &
   \left(\frac{m\omega}{\pi}\right)^{1/4} 
   \frac{1}{\sqrt{24}(m\omega)^{5/2}}
   e^{-\frac{p^2}{2m\omega}} \\
   & (3(m\omega)^2 - 12 p^2 m\omega +4p^4)
 \end{split}
\label{}
\end{equation}

Use $E_n = (n+1/2) \omega$, and substitute all the above relations
into the first order perturbation with the first two even terms, we have:
\begin{equation}
 \begin{split}
   \braket{p}{\varphi} =& \left(\frac{1}{m\omega\pi}\right)^{1/4} 
   e^{-\frac{p^2}{2m\omega}}\\
   &\Big\{ 1 - \frac{A(\alpha-1)^2}{8\omega\alpha^{5/2}}
   \left(\frac{p^2}{m\omega}-p_1\right)\left( \frac{p^2}{m\omega}-p_2
   \right) \Big\}.
 \end{split}
\label{eqn:phiformula}
\end{equation}

Here $p_1$ and $p_2$ are defined as:
\begin{align}
p_1 & = \frac{\sqrt{6\alpha^2+4\alpha+6}-(\alpha+3)}{2(\alpha-1)},\\
p_2 & = \frac{-\sqrt{6\alpha^2+4\alpha+6}-(\alpha+3)}{2(\alpha-1)}.
\label{}
\end{align}
Since $\alpha>1$ by definition, the relations above are always well defined.
Also note that:
\begin{equation}
6\alpha^2+4\alpha+6-(\alpha+3)^2 = (\alpha-1)(5\alpha+3)>0,
\label{}
\end{equation}
and it holds that:
\begin{equation}
p_1>0, \qquad p_2 < 0.
\label{}
\end{equation}

The potential barrier corresponds to the case when $A>0$. No matter how
small $A$ is, there always exists a $p^2/(m\omega)>p_1$ and
Equation~\eqref{eqn:phiformula} equals to zero. The
position of the node can be easily calculated from
Equation~\eqref{eqn:phiformula}. For instance, for
$A=0.004$ (the low barrier case), Equation~\eqref{eqn:phiformula} gives the node position at $p=\pm 12.99$. Compared to Fig.~\ref{fig:lowb},
Equation \eqref{eqn:phiformula} gives a good prediction of the position of the
extra node.

It is also of interest to look at the case of $A<0$, that is, when the potential
barrier is substituted by a potential well. The factor inside the braces in
Equation~\eqref{eqn:phiformula} reaches its minimum at $p=0$, and this
minimum is
\begin{equation}
1-\frac{A(\alpha-1)^2}{8\alpha^{5/2}\omega} p_1 p_2 
\label{}
\end{equation}
Note that $p_1p_2$ is a fixed number and only depends on $\alpha$,
therefore when $A$ is small, this minimum is always positive. Then it
follows that $\braket{p}{\varphi}>0$ everywhere, and there is no node.
For example, when $A=-0.004$, the minimum is $0.94$ and far from zero.
\end{appendix}



\begin{thebibliography}{10}
\bibitem{nodenote}
Note that at finite temperature, what we refer to as a ``node'' is not,
  strictly speaking, an exact zero.
\end{thebibliography}

\end{document}